\documentclass{kapproc} 

\usepackage{procps} 

\usepackage[dvips]{graphicx}

\upperandlowercase

\kluwerbib 

\def\gtaprx{\mathrel{\vcenter{\offinterlineskip \hbox{$>$}
    \kern 0.3ex \hbox{$\sim$}}}}

\begin{document}

\articletitle[Quantifying Bar Strength]
{Quantifying Bar Strength:\\
Morphology Meets Methodology}

\author{Paul B.~Eskridge\altaffilmark{1}}
 
\affil{\altaffilmark{1}Department of Physics \& Astronomy, Minnesota State
University, Mankato MN 56001, USA}

\begin{abstract}
A set of objective bar-classification methods have been applied to the Ohio
State Bright Spiral Galaxy Survey (Eskridge et al.~2002).  Bivariate 
comparisons between methods show that all methods agree in a statistical sense. 
Thus the distribution of bar strengths in a sample of galaxies can be robustly 
determined.  There are very substantial outliers in all bivariate comparisons.
Examination of the outliers reveals that the scatter in the bivariate
comparisons correlates with galaxy morphology.  Thus multiple measures of bar 
strength provide a means of studying the range of physical properties of galaxy 
bars in an objective statistical sense.
\end{abstract}

\begin{keywords}
galaxies: fundamental parameters --- galaxies: spiral --- galaxies: statistics 
--- galaxies: structure --- infrared: galaxies
\end{keywords}

\section{Introduction}

The existence and importance of bars in spiral galaxies has been recognized for 
the entire history of extragalactic astronomy (Curtis 1912).  Hubble (1936) 
included the barred (SB) and unbarred (S) classes as the tines of the tuning 
fork.  de Vaucouleurs (1959) first noted that ``barredness'' is not a binary 
state, and invented the class SAB for galaxies that are detectably, but not 
strongly barred.  The relevance of observed wavelength to the detection of bars 
was first pointed out by Hackwell \& Schweizer (1983) who discovered a strong
bar in their $H$-band image of the optical SAB NGC 1566.  The most extensive 
demonstration of this issue to date is that of Eskridge et al.~(2000).  One of 
the shortcomings of the above-cited work is that it uses subjective bar 
classification.  Subjective classification is a useful starting point, but to 
move toward a physical understanding of morphology requires the general
adoption of an appropriate, quantifiable and objective metric.  The problem is 
that bars do not lend themselves naturally to scalar quantification.  There 
have been a large number of ``bar-strength quantifiers'' proposed and 
implemented by different groups, and all of them measure somewhat different 
things (e.g., Elmegreen \& Elmegreen 1985; Ohta et al.~1990; Martin 1995; 
Wozniak et al.~1995; Abraham \& Merrifield 2000; Buta \& Block 2001).  

This presentation reports on the progress of a study to intercompare a set of
different bar-strength quantifiers using a single large, and statistically
complete sample of nearby spiral galaxies (the OSU sample -- see Eskridge et 
al.~2002).  The main purposes of the study are two-fold:

{\noindent 1) To determine how well different bar-strength classifiers compare
to one another and study the sorts of situations in which they fail to agree.}

{\noindent 2) To address how reliably we can study the bar fraction of high-$z$
galaxies.  This follows from a comparison of the results from techniques that 
are most demanding of the observational data with those that are the least 
demanding.}

Section 2 presents the methods included in this study.  Section 3 presents some 
of the bivariate relations found.  Section 4 includes a discussion of the 
findings to date and a summary of issues yet to be fully investigated.

\section{Summary of Methods}

The most data intensive method included in this study is the $Q_g$ 
``disk-torque'' measure of Buta et al.~(2004).  It is a refinement of the $Q_b$ 
``bar-torque'' measure of by Buta \& Block (2001).  Both of these methods are 
pure NIR methods by definition.  The goal of these methods is to express the 
strength of the bar in terms of its gravitational effect on the surrounding 
disk.  In principle, this is an excellent physical means of measuring bar 
strength.  In practice, the methods requires assumptions about disk structure 
that are difficult to verify.  They also require high spatial resolution, high 
S/N images, and are very laborious to implement.

Both of these methods are based on Fourier-mode decomposition.  In its simplest 
guise, the Fourier-mode method amounts to measuring the strength of the
constant-phase $m=2$ mode in an image.  This study uses analysis by
D.~Elmegreen and her students, following Elmegreen \& Elmegreen (1985).

A method proposed by Abraham \& Merrifield (2000) measures bar strength by the 
maximum ellipticity of the inclination-corrected disk.  It is thus most 
sensitive to long, thin bars.  The method was refined by Whyte et al.~(2002),
who also applied it to the OSU survey sample.  This study uses the results of 
Whyte et al.~(2002).

The complete study will consider the neural-net method of Odewahn et 
al.~(2002), developed for automated classification of high-redshift galaxies.  
It is the most forgiving of data quality.  These results are not yet available.

\section{A Selection of Bivariate Results}

Figures 1--3 show a selection of bivariate comparisons amongst $Q_g$ (Buta et 
al.~2004), the $H$-band m$=$2 Fourier mode amplitudes (from D.~Elmegreen), and 
the $B$- and $H$-band Abraham bar parameter $F_{bar}$ (Whyte et al.~2002). In
all cases, statistically significant correlations exists (at better than the
$10^{-4}$ level from a variety of tests), but with substantial scatter and 
large outliers.  The dashed lines are the ordinary least-squares bisectors.

\begin{figure}[ht]
\centerline{\includegraphics[width=4in]{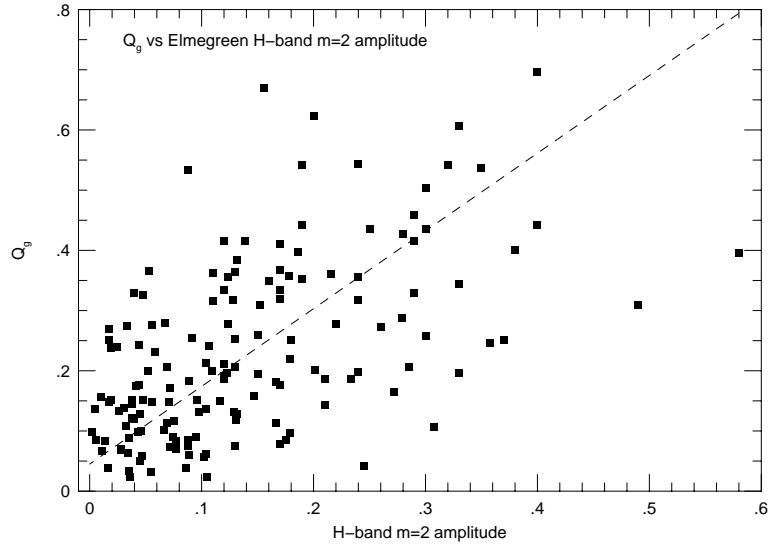}}
\caption{$H$-band $m=2$ Fourier mode against $Q_g$.}
\end{figure}

\begin{figure}[ht]
\centerline{\includegraphics[width=4in]{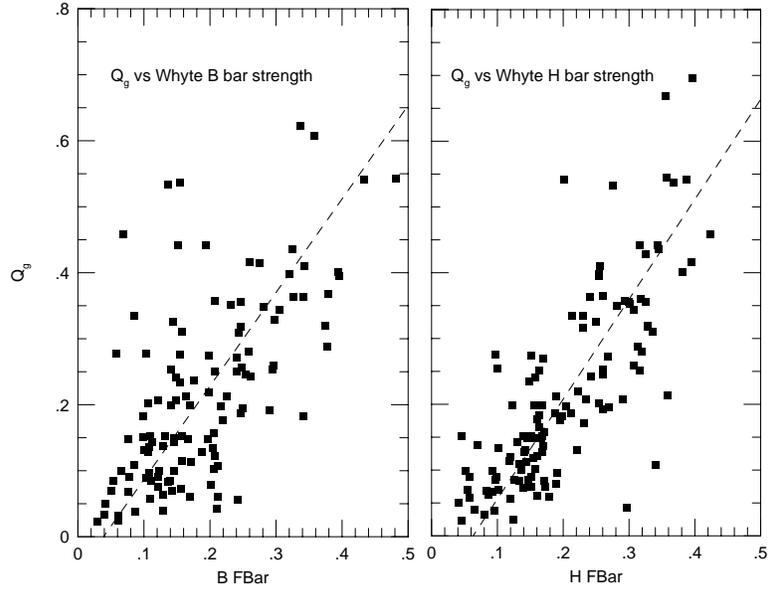}}
\caption{$Q_g$ vs a) $H$-band $F_{bar}$ and b) $B$-band $F_{bar}$.}
\end{figure}

\begin{figure}[ht]
\centerline{\includegraphics[width=4in]{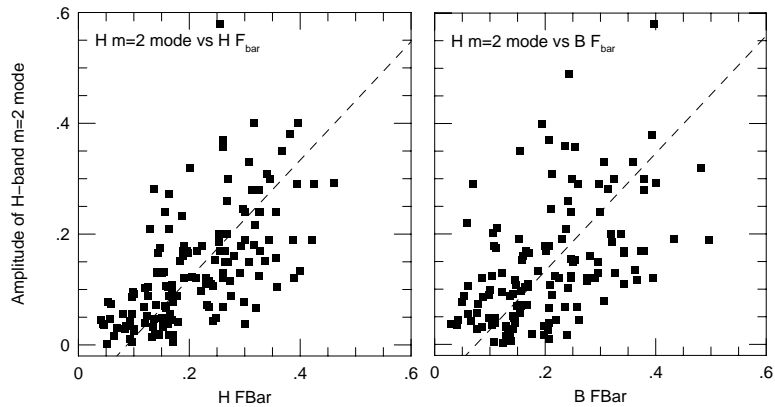}}
\caption{$H$-band Fourier $m=2$ mode vs a) $H$-band $F_{bar}$ and b) $B$-band 
$F_{bar}$.}
\end{figure}

\section{Summary and Future Work}

It is a great reassurance that all the methods examined here provide 
statistically self-consistent results, even when comparing different methods in 
different wave-bands (NIR torque methods compared with $B$-band bar 
ellipticity, as an example).  Thus any of the methods discussed here can 
provide a good statistical means of measuring bar strength.  This argues that 
bar-strength analysis of high-redshift galaxy samples can, at least in 
principle, be compared with analysis of nearby systems even when the
observations are different rest-frame wavelengths, and at very different 
spatial sampling.  The inclusion of neural-net techniques to the current study 
promises to be a very interesting advance.

A more careful examination of the bivariate results reveals that even very 
similar methods tend to have significant scatter between them, and the 
occasional very substantial outlier.  The existence of these outliers actually 
gives us the opportunity to study the failure modes of the various techniques 
under consideration.  As an example, the two systems with the largest $H$-band 
$F_{bar}$ for their $Q_g$ values (the points on the bottom right of Fig.~2b) 
are NGC 3166 and NGC 4772.  Both these galaxies have $H$-band morphologies of 
SB0 (Eskridge et al.~2002), with relatively large bulges and weak disks.  The
$F_{bar}$ statistic is very good at describing such bars, but $Q_g$ is not.
This is due to the relatively strong bulges causing the effective torque from
the bars to be small.  In the same plot, the two galaxies with the largest 
$Q_g$ for their $H$-band $F_{bar}$ values are NGC 1042 and NGC 3319.  Both 
these systems have grand design spiral patterns and short bars.  The shortness
of the bars appears to drive the low $F_{bar}$ values.  But the combination of
the bars and the strong spiral patterns results in a large disk torque.  A true 
$Q_b$ analysis, as discussed in Buta et al.~(2003), will improve the 
sensitivity of torque-based analysis to pure bar components.  The scatter in 
the bivariate plots is generally correlated with morphological patterns in the 
sample.  The use of multiple measures of bar strength thus offers us the 
opportunity to study full range of the physical properties of bars in an 
objective, statistical sense.  

\begin{acknowledgments}
This study is made possible by the work of the authors of Block et al.~(2002), 
Buta et al.~(2004) and Whyte et al.~(2002), and D.~Elmegreen and her students.  
I thank them for their efforts.  Thanks are also due to David Block and Ken 
Freeman for inviting me, and for organizing the conference.  The OSU survey was
made possible by grants AST-9217716 and AST-9617006 from the US National 
Science Foundation.  This work is supported by Minnesota State University and 
by NASA grant HST-GO-09892.02.
\end{acknowledgments}

\end{document}